\documentclass[sn-mathphys,Numbered]{sn-jnl}
\usepackage{graphicx}%
\usepackage{multirow}%
\usepackage{amsmath,amssymb,amsfonts}%
\usepackage{amsthm}%
\usepackage{wasysym}%
\usepackage{mathrsfs}%
\usepackage[title]{appendix}%
\usepackage[dvipsnames]{xcolor}%
\usepackage[normalem]{ulem}
\usepackage{multirow}
\usepackage{geometry}
\newgeometry{vmargin={20mm}, hmargin={26mm,26mm}, columnsep={10mm}}

\usepackage{lineno}

\begin{document}


\title[Article Title]{ 
Measuring high-precision luminosity at the CEPC }

\author[1]{ \fnm{Jiading}   \sur{Gong}}
\author[2]{ \fnm{Jun}       \sur{He}}
\author[2]{ \fnm{Rongyan}   \sur{He}}
\author*[3]{\fnm{Suen}    \sur{Hou}}  \email{suen@as.edu.tw}  
\author[2]{ \fnm{Quan}   \sur{Ji}}
\author[4]{ \fnm{Renjie} \sur{Ma}} 
\author[4]{ \fnm{Ming}     \sur{Qi}} 
\author[2]{ \fnm{Haoyu}    \sur{Shi}} 
\author[1]{ \fnm{Weimin}       \sur{Song}} 
\author[4]{ \fnm{Xingyang}     \sur{Sun}} 
\author[2]{ \fnm{Haijing}  \sur{Wang}}
\author[4]{ \fnm{Yilun}    \sur{Wang}} 
\author[4]{ \fnm{Jialiang}    \sur{Zhang}} 
\author[4]{ \fnm{Lei}         \sur{Zhang}} 

\affil[1]{\orgdiv{Jilin University}, 
\orgaddress{\city{Changchun},   \postcode{130012}, \country{China}}}

\affil[2]{\orgdiv{Institute of High Energy Physics},  
\orgaddress{\city{Beijing},  \postcode{100049}, \country{China}}}

\affil[3]{\orgdiv{Academia Sinica},
\orgaddress{\city{Taipei}, \postcode{115201}, \country{China}}}

\affil[4]{\orgdiv{Nanjing University}, 
\orgaddress{\city{Nanjing},  \postcode{210093}, \country{China}}}

\vspace{-10mm}
\abstract{

\textbf{Purpose:} 
Luminosity measurement at the Circular Electron-Positron Collider (CEPC)
is required to achieve 10$^\textmd{-4}$ precision when operating
at the center-of-mass energy of the Z-pole.
Approximately 10$^\textmd{12}$ Z-bosons will be collected  
to refine measurements of Standard Model processes. 
The design of the luminosity calorimeter (LumiCal)  takes into account the geometry
of the Machine-Detector-Interface (MDI) 
for detection of Bhabha events.
The detector simulation with GEANT predicts measurements 
of scattered electrons, positrons, and radiation photons.
\\

\textbf{Methods:}
Bhabha events are generated using the BHLUMI program. 
Electrons arriving within the LumiCal fiducial coverage are counted. 
{\color{black} The scattering angles of electrons are measured}, 
taking into account deviations caused by multiple scattering and detector resolution.
 {\color{black} The error of the mean for events passing}
the lower fiducial edge in polar angle, $\theta_{acc}$, shall be better than 1~$\mu$rad,
which corresponds to 10$^\textmd{-4}$ on the Bhabha cross section. 
{\color{black} In each beam direction,}
the LumiCal design features two silicon position detectors to measure
electron impact positions, and $ 2\, \textit{X}_0$ LYSO bars behind for 
$e^\pm/\gamma$ identification.
The $13\, \textit{X}_0$ LYSO calorimeter, mounted 
{\color{black} further downstream} in front of the quadrupole magnet, 
is used to measure energy of electrons.
\\

\textbf{Results:} 
The luminosity measurement by counting Bhabha events 
depends on the accuracy of
 {\color{black} the lower acceptance angle ($\theta_{acc}$) 
 at the detector's fiducial edge. 
}
The beam-pipe design incorporates low-mass windows of 
{\color{black} 
1~mm thick beryllium (Be) layers
\color{black} to reduce multiple scattering effects.  
The LumiCal has pixelated silicon detectors with better than 5 $\mu$m resolution  
and LYSO crystals segmented into $3{\times} 3$~mm$^2$, which enhances the  
capability for detecting radiative Bhabha events.}
To achieve a precision level of $10^{-4}$,
it is crucial to monitor the interaction point (IP) of 
colliding beams and the positions of detectors 
with the error on mean of {\color{black} better than 1~$\mu$rad.}
\\

\textbf{Conclusion:}
{\color{black}
The LumiCal measures Bhabha scattering events using 
Si-detectors and finely segmented LYSO arrays.
Its design is optimized for detecting radiative photons 
that are separated from electrons by a sufficiently large opening angle. 
This measurement aims to detect higher-order corrections 
to the Bhabha interaction. 
Emphasis is placed on steering the beams for IP distribution   
and survey monitoring of detector positions to achieve 
high-precision luminosity measurements. }
}


\maketitle
\twocolumn

\setcounter{section}{1}
\renewcommand*\thesubsection{\arabic{subsection}}

\section*{Introduction}
\label{sec:lumi_intro}

The Circular Electron Positron Collider (CEPC) is proposed 
as a Higgs factory, targeting an instant luminosity of 
$5 \,{\times}\, 10^{34}/\rm{cm}^{2}$s$^{1}$ at a center-of-mass energy of 240 GeV 
{\color{black} for ten years, to produce 2.6 million HZ events.
The Z-pole operation will follow for two years 
with the instant luminosity of 
 $1.15 \,{\times}\, 10^{36}$/cm$^{2}$s$^{1}$ to 
accumulate $2.5 \,{\times}\, 10^{12}$ Z-boson events \cite{CEPC_TDR}.}
With such high event statistics, the measurement of Standard Model processes
will require a reference luminosity of better than $10^{-4}$ in precision.

 
{\color{black}
The CEPC colliding beams are designed with a 33~mrad crossing angle
at the Interaction Point (IP).
The beam pipe has a 20~mm diameter at the IP, 
which expands horizontally into an oval \color{black} 
  race-track shape into two separated beam pipes on the horizontal 
  $x$-$z$ plane.
  The $z$-axis of the laboratory frame
  is the bisector between the two beam axes.
  The $x$-axis is pointing toward the accelerator ring center,
  and the $y$-axis is vertical upwards. The colliding $e^+$, 
  $e^-$ beams are tilting at $16.5$~mrad to the $z$-axis.
} 
The Luminosity Calorimeter (LumiCal) 
is proposed at the positions
{\color{black} in front of} the inner beam-pipe flanges 
at $|z|$ = 670~mm, {\color{black}
  and the region} 
{\color{black} of $|z|$ = 780~mm to 1100~mm} 
between the bellows and the quadrupole magnets.
The polar angle coverage is limited to less than 85 mrad.

The luminosity of $e^+ e^-$ colliding beams 
{\color{black} can be accurately}
 measured using the elastic scattering events known as the Bhabha interaction.
The LumiCal will detect 
back-to-back electron and positron pairs,
each with  
{\color{black} energy}
 equal to the beam energy.
In the center-of-mass (CM) frame, the leading-order Bhabha cross-section is expressed as
\begin{equation}
 \sigma = \frac{16 \pi \alpha^2}{ s} \left(
          \frac{1}{\theta_{min}^2} - \frac{1}{\theta_{max}^2} \right),
\label{eq:BabaSigma} 
\end{equation}
{\color{black}
for the polar angle range of ($\theta_{min} \,{<}\, \theta \,{<}\, \theta_{max}$) 
with the azimuthal angle integrated. 
}

The integrated luminosity $\mathcal{L}$ of $e^+ e^-$ collisions,
for the number of $N_{acc}$ Bhabha events detected, is given by
\begin{equation}
          \mathcal{L} = \frac{1 }{ \epsilon }  
          \frac{ N_{acc}}{\sigma }.
\label{eq:dLL} 
\end{equation}
\noindent
where $\epsilon$ is the detection efficiency.
\noindent
The error in luminosity is derived from Eq.~\ref{eq:BabaSigma}
to be
\begin{equation}
          \frac{ \Delta \mathcal{L}}{ \mathcal{L} } 
          \sim  \frac{ 2\Delta \theta} { \theta_{min} },
\label{eq:dLL} 
\end{equation}
{\color{black}
in which $\theta_{max} \,{\gg}\, \theta_{min}$ is assumed.}  
The systematic uncertainties primarily arise from 
{\color{black} the position offset of the detector acceptance 
with $\Delta\theta$ at $\theta_{min}$.
}
The error on luminosity is twice the error on
$\theta$ at the lower edge of the $\theta$ acceptance.
Assuming a $\theta_{min}$ of 20~mrad, a deviation 
corresponding to a $10^{-4}$ error on luminosity is 
$\Delta\theta$ = 1~$\mu$rad.
At a distance of $|z|$ = 1~m from the $e^+ e^-$ interaction point,
the tolerance for the survey of the detector inner edge 
in the transverse direction is 1~$\mu$m.

{\color{black}
The LumiCal geometry is optimized for acceptance 
in the laboratory frame. The detector fiducial edge
does not correspond to a constant $\theta_{min}$ in the CM frame.
}
In the following, we first present 
{\color{black} 
  in section {\it ``LumiCal design''} the detector modules 
  in the Machine Detector Interface (MDI) region.
}
The Bhabha interaction of $e^+ e^- \to e^+ e^- (n\gamma)$ 
is simulated with the
BHLUMI event generator \cite{Jadach:1996is}. 
The LumiCal acceptance  is evaluated and
discussed in the {\it ``Bhabha event acceptance''} section.  

The generation of radiative photons in BHLUMI employs the 
Yennie-Frautschi-Suura (YFS) exponentiation algorithm~\cite{Jadach:1991by,YFS} 
to combine contributions of real and virtual radiation photons,
with cancellation of infrared divergent terms to all orders in $\alpha$.
{\color{black}
The detection of radiative photons helps to validate the 
higher-order corrections to the Bhabha scattering
interaction.}
In the section {\it ``Detecting radiative Bhabha''},
photons near scattered electrons are simulated. 
The methods and results for the detection of radiative photons are discussed.

The LumiCal performance for detecting Bhabha events 
{\color{black} is simulated with the GEANT program \cite{GEANT3,GEANT4}.
The results are discussed in the {\it ``LumiCal simulation''} section.
The impact positions of electrons on the Si wafers }
and the electromagnetic (EM) shower distributions in LYSO crystals 
are evaluated for event acceptance and systematic errors.
The upstream beam pipe material causes multiple scattering
and deviations in the electron impact positions. The effects are estimated.

{\color{black} 
Photons traversing Si-wafers do not generate ionizing charges
for detection.
In combination with the LYSO arrays, which detect EM
showers of both electrons and photons, the $e/\gamma$ identification
can thus be conducted.
The analysis method is discussed in the 
{\it ``LYSO crystals before flanges''} section.
Bhabha events are identified with electrons carrying beam energy.
The long LYSO bars are implemented for this purpose. 
The simulation of energy deposits is discussed in the 
{\it ``LYSO crystals behind Bellows''} section.   
}

{\color{black} 
Bhabha scattered electrons 
are measured relative to the IP of colliding beams. \color{black}
The distribution of IP is monitored with the beam currents 
measured by}
Beam Position Monitors (BPMs), in positions inside the 
beam-pipe flanges and before the quadrupole magnets surrounding each beam pipe.
The precision in measuring electron scattering angle is
discussed in the section
{\it ``Precision on LumiCal $\theta$ acceptance''}, 
for the IP distribution and the survey and monitoring of LumiCal.
The challenge in achieving a high precision luminosity 
measurements is summarized in the {\it ``Summary''} section.

\begin{figure}[b!]  
  \centering                         
  \includegraphics[width=.95\linewidth]{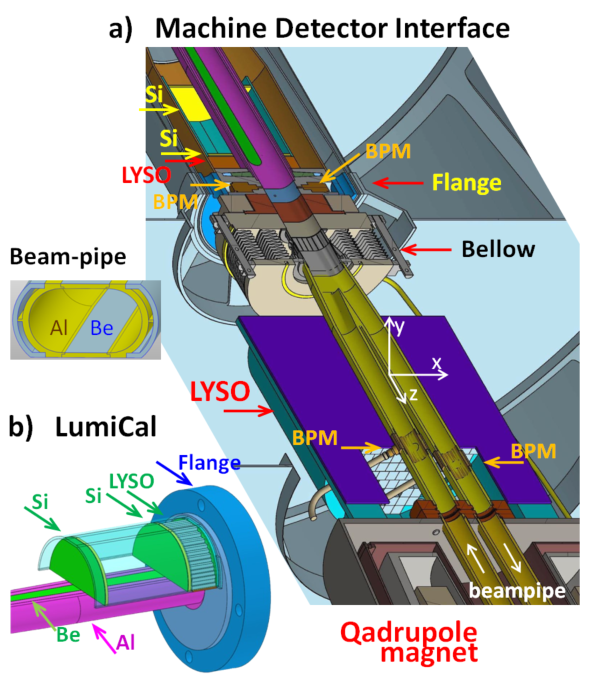}
  
  
  \caption{  \label{fig:MDI-LumiCal3D}
  a) The cut-view on one side of the IP is shown
  for the MDI region containing the LumiCal modules mounted before
  the flange of the race-track beam-pipe and 
   {\color{black} 
   in the region between the bellows and quadrupole magnets}.
  b) The LumiCal modules located before the flange consist of two Si-wafers 
  and $2\, X_0$
  LYSO bars arranged above and below the beam pipe. 
  The green-colored Be layers serve as low-mass windows in conjunction with 
  dual-layer Al pipes for circulation of cooling water.
  }
\end{figure}  

\section*{LumiCal design  }
\label{sec:LumiCal}

The LumiCal detector in the MDI region
{\color{black} 
is illustrated in Fig.~\ref{fig:MDI-LumiCal3D}.a.
The inner beam pipe of $\diameter 20$~mm at IP is terminated 
with the flanges at $|z|$ = 700~mm, to be joined with the
bellows. The beam line branches into two pipes before the quadrupole.  
The beam pipe design  
is optimized for the least heating loss due to higher-order modes \cite{CEPC_TDR}.
}
The flat surfaces of the beam pipe are {\color{black} equipped with}
a low-mass beryllium (Be) window of 1~mm thick 
between the aluminum (Al) dual-layer pipes.

{\color{black}
On each side of the IP, the LumiCal has two layers of Si-detectors and LYSO crystal 
arrays of $2\, X_0$ (23 mm), positioned in front of} the flanges
 (Fig.~\ref{fig:MDI-LumiCal3D}.b). 
The Si-wafers are located at $|z|$ = 560 and 640~mm.
{\color{black}
The LYSO crystals are segmented into $3{\times} 3$~mm$^2$ arrays on the $x$-$y$ plane 
with the front surface located at $|z|$ = 647~mm.       
}

Behind the flanges and bellows of a total $4.3\, X_0$ of steel, 
the second sections of long LYSO arrays of $13\, X_0$ (150 mm) 
are mounted in front of the quadrupole magnets.
{\color{black}
These LYSO arrays are segmented in $10 \,{\times}\, 10$~mm$^2$ to
cover the Molière radius of electromagnetic showers 
in $3 \,{\times}\, 3$ arrays for electrons carrying beam energy.
}

\begin{figure*}[t!]  
  \centering
  
  \includegraphics[height=.30\linewidth]{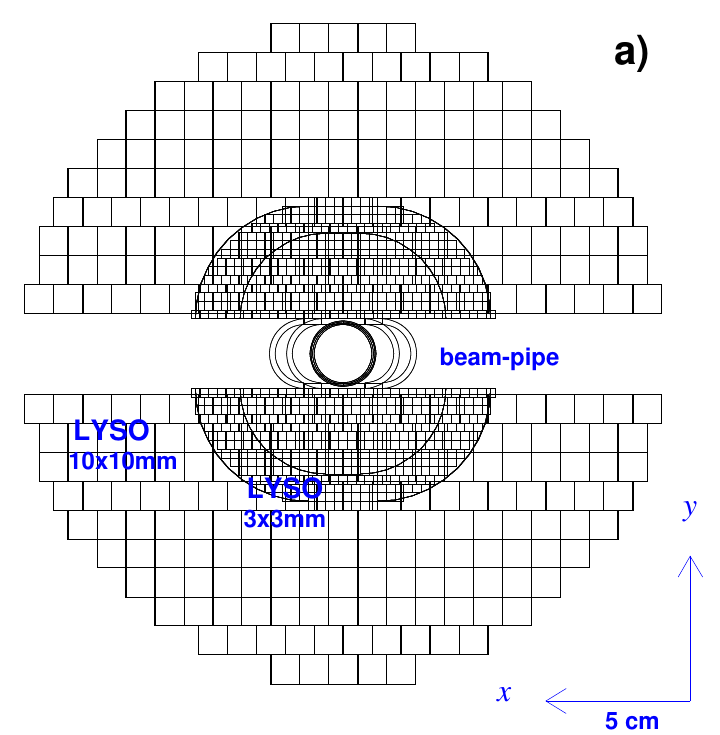}
  \hspace{8mm}
  \includegraphics[height=.30\linewidth]{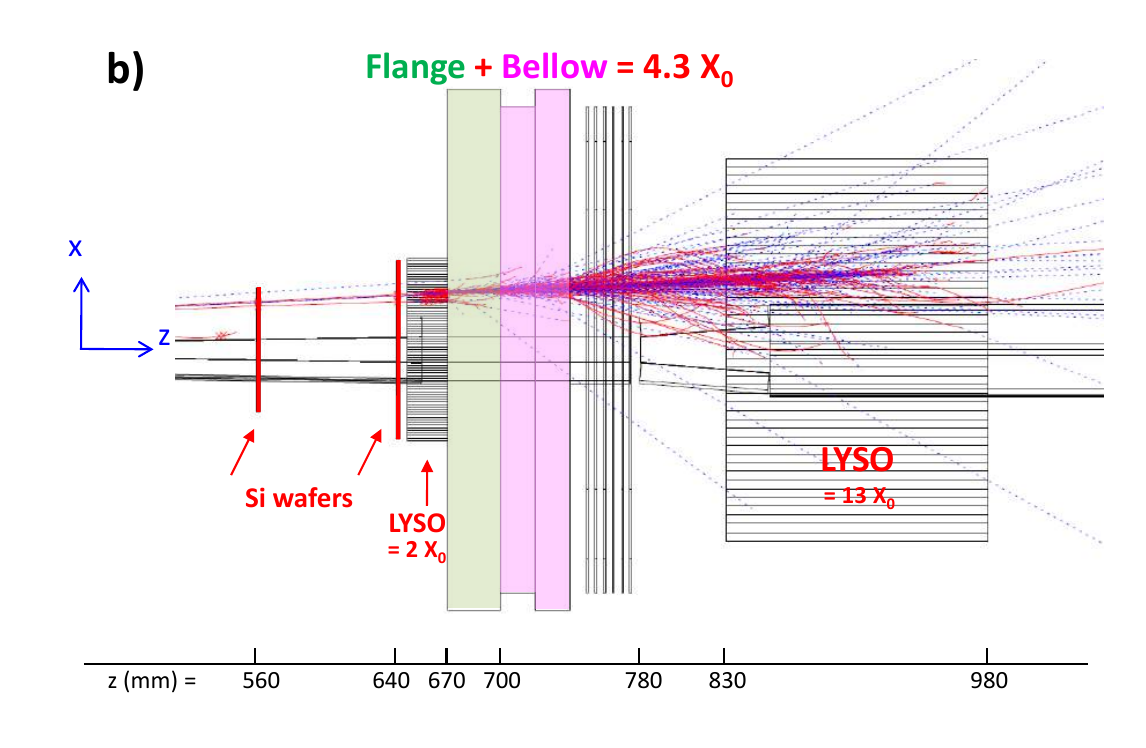}
  
  \caption{ \label{fig:LumiCal_flange}
  The LumiCal design in the GEANT simulation is depicted in two views:
  a) the $x$-$y$ perspective of the LYSO bars in front of the flange 
  and behind it,  with dimensions of
  $3{\times} 3$~mm$^2$ and $10{\times} 10$~mm$^2$, respectively,
  positioned above and below the race-track pipe;
  b) the $x$-$z$ view of a 50 GeV electron shower in the Si-wafers 
  and LYSO crystals, in front and behind the flange and bellow of 
  the beam-pipe facility.   
  }
\end{figure*}

{\color{black}   
 The precision of LumiCal acceptance for Bhabha events 
 is most affected by the offset of the lower edge of Si-detectors
 used to measure the electron impact positions.
}
The active area of the front Si-wafer is positioned perpendicularly
with $|y|\,{>}\, 12$~mm, which corresponds to 
$\theta_{lab} \,{>}$ $\mathrm{atan} (12/560)$, to be larger than 21.4~mrad.
The LumiCal fiducial area is symmetric {\color{black}
in both $x$ and $y$ around the $z$-axis
\color{black}  within the range of
 ``$25 \,{<}\, \theta^{b.p.\pm}_{acc} \,{<}\, 80$~mrad'', 
 {\color{black} measured to the beam directions}. 
 This includes the semicircles to the closest beam pipe centers,
 and the joint rectangular area in between.  
}

The beam-crossing of 33 mrad has the effect of boosting colliding 
particles toward the $+x$ direction.  
{\color{black}
 The scattered $e^+$ and $e^-$ of Bhabha interactions are 
 distributed symmetrically around the outgoing beam directions.
 When projected on the $x$-$y$ plane at $|z|$, 
 the beam centers are located at $x= \pm \tan(0.0165) {\times} |z|$~mm and $y$ = 0.
}

{\color{black}
 The region within the height of the beam pipes ($|y|\,{<}\, 12$ mm) has 
 a wider span in $x$ taken by the beam pipe. In this region,
 the Bhabha event acceptance is small for both $e^+$ and $e^-$ being detected.
 This area is spared from the detector coverage.}

{\color{black}
Si-detectors are chosen for measuring the impact positions of
charged electrons.
The well-developed {\color{black} double-sided}
strip detectors, fabricated on 300~$\mu$m thick 
wafers in 50~$\mu$m pitches, are the baseline option.
  {\color{black}
  The strips oriented in $x$ and $y$ directions
  can provide a uniformly pixelated coverage.}
{\color{black}
Ionizing charges generated by  
a traversing electron (80 e-hole pairs per $\mu$m)
are collected in adjacent strips.}
The spatial resolution utilizing charge-sharing 
characteristics can reach 5~$\mu$m \cite{Si-PDG,Hou}.}
{\color{black} With the measurements of two Si detectors 
matching the EM shower in LYSO arrays,
the detection efficiency for electrons 
can be estimated.}

{\color{black}
 The compactness of readout electronics for the finely segmented
 LYSO bars can be achieved with millimeter-sized 
 silicon photomultipliers (SiPMs). 
 With the Si-detectors positioned in front, which measure
 charged particles, the
 identification of $e/ \gamma$ is facilitated.   
}

The LumiCal design has been simulated using the GEANT program.
The layout of LYSO bars in the $x$-$y$ view is illustrated in 
Fig.~\ref{fig:LumiCal_flange}.a.
The event display of a 50 GeV electron shower is plotted
in Fig.~\ref{fig:LumiCal_flange}.b,
in the $x$-$z$ view, which shows the particle tracks in the 
LumiCal detector modules.

\begin{figure}[b!] 
  \centering       
  \vspace{-4mm}
  \includegraphics[width=0.95\linewidth]
  {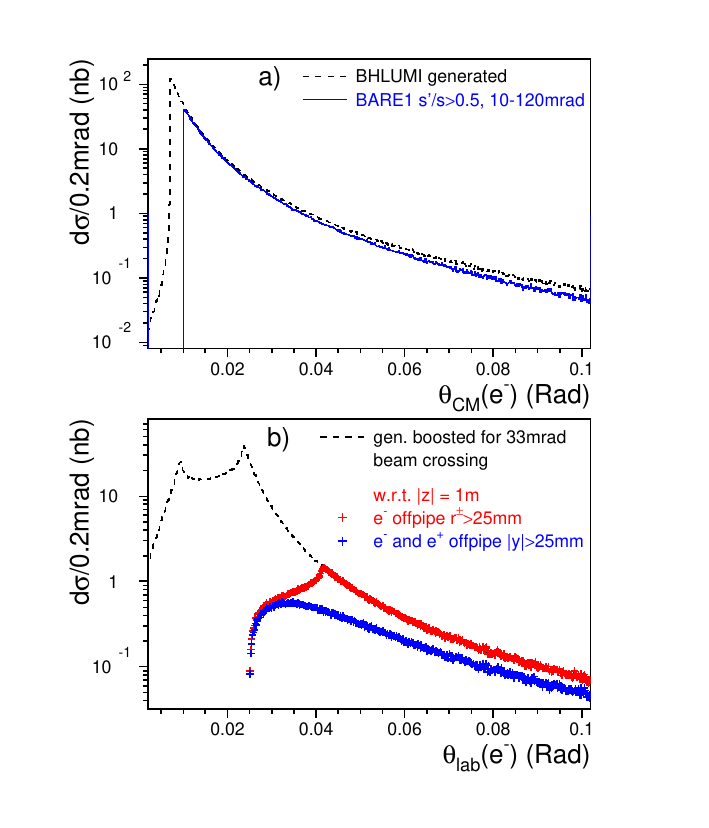}
  
  \vspace{-6mm}
  \caption{ \label{fig:bhlumi-th-b2b}   
  \color{black}
  BHLUMI scattering electrons are plotted for distributions of a) electron polar angle 
  in the CM frame, for the generated and selected in the $\theta_{\rm CM}$
  range specified {\color{black}
    (10 to 120 mrad, selected with BARE1 and 
    $s'(e^+,e^-)/s\,{>}\, 0.5$ \cite{Jadach:1996is}),} and 
  b) boosted for beam-crossing in the laboratory frame (black-dashed).
    The distribution of electrons in the LumiCal acceptance 
  projected at $|z|$ = 1~m, are plotted for {\color{black}
  electrons with {\color{black} radius}
  $r^\pm\,{>}\, 25$~mm (red marks, distance to $x$-axis between the beam centers 
  of $x$ = $\pm16.5$~mm), and for
  both $e^+$ and $e^-$ of $|y|\,{>}\, 25$~mm (blue marks).} 
  }
\end{figure}  

\begin{figure}[b!] 
  \centering
  \includegraphics[width=.95\linewidth]
  {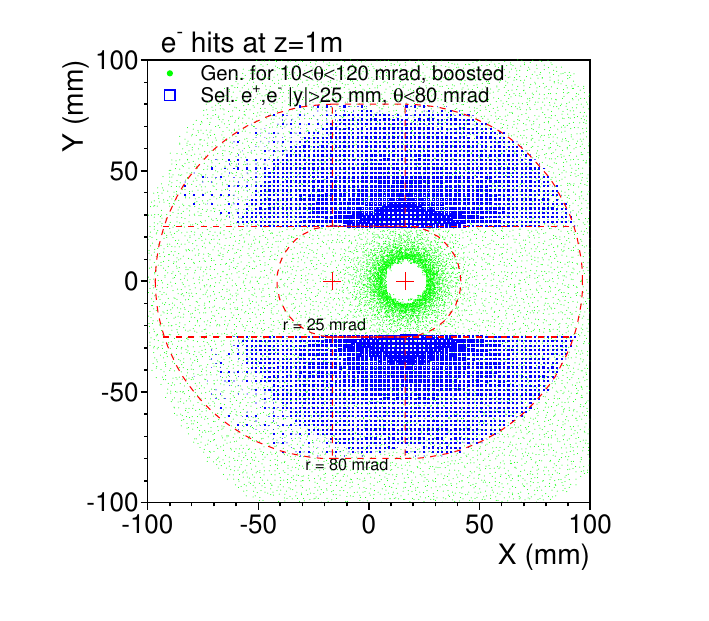}
  \vspace{-6mm}
  \caption{ \label{fig:bhlumi-b2bxy}   
  The green points are electron hits generated using BHLUMI 
  ($\theta_{\rm CM}$ in $10-120$ mrad), which are boosted
  and projected at $|z|$ = 1~m. 
  The blue boxes are hits with $e^+$ and $e^-$ 
  selected in the LumiCal fiducial region (in red dashed lines), 
  for the semicircular area of $25 \,{<}\, \theta^{b.p.\pm}_{acc}\,{<}\, 80$~mrad to 
  the closest beam pipe centers and the joint rectangles in between,
  excluding the areas below the beam pipe of $|y|\,{>}\, 25$~mm.
  }
\end{figure} 

\section*{Bhabha event acceptance}
\label{sec:Xsection}

Bhabha scattering events are simulated using the BHLUMI at
 $\sqrt{s}$ = 92.3~GeV with the default parameters of the program 
 (in demo.f \cite{Jadach:1996is}). {\color{black}
The polar angle in the center-of-mass frame, $\theta_{\rm CM}$},
 is specified in the range of 10 to 120~mrad.
The program produces events in a wider range 
($7 \,{<}\, \theta_{\rm CM} \,{<}\, 240$ mrad). 
{\color{black}
 In Fig.~\ref{fig:bhlumi-th-b2b}.a, the generated and 
 selected events are plotted. 
}

At CEPC, the final state particles are boosted and are
plotted in Fig.~\ref{fig:bhlumi-th-b2b}.b for the $\theta$ distributions 
of electrons in the laboratory (lab) frame.  
The maximum shift of $\pm 16.5$~mrad occurs
for electrons traveling along the $x$-axis.
The distributions are also plotted for electrons selected in the LumiCal
acceptance projected at $|z|$ = 1~m,
with {\color{OliveGreen} radius}
$r^\pm \,{>}\, 25$~mm to the beam pipe centers (red marks), and for 
Bhabha events selected with both {\color{OliveGreen} $e^+$ and $e^-$}
in $|y|\,{>}\, 25$~mm (blue marks).

The distributions of Bhabha scattering electrons in the $x$-$y$ plane 
(at $|z|$ = 1~m)
are plotted in  Fig.~\ref{fig:bhlumi-b2bxy}.
The dashed line illustrates the fiducial range of the LumiCal. 
The boosted electrons are distributed
around the $+x$ beam center (green points).
The distribution in blue boxes illustrates events with both $e^+$ and $e^-$ detected
in the LumiCal coverage.

The cross-section of Bhabha scattering events within the 
LumiCal fiducial acceptance has been estimated with the BHLUMI 
for the selection criteria with 
one or both back-to-back $e^+$ and $e^-$ detected.
Listed in Table~\ref{tab:bhlumi_xsection} are the results obtained for  
$\sqrt{s}$ = 92.3~GeV,
with two $\theta^{b.p.\pm}_{acc}$ thresholds of 20 and 25 mrad.
The lower threshold of 20 mrad yields a cross-section that is nearly double 
that of the 25 mrad. 
{\color{black}
At the sides of the beam pipe ($|y|\,{<}$ 20, 25~mm, at $|z|$ = 1~m),
the $\theta^{b.p.\pm}_{acc}$ is widened by 33 mrad, and thus
the events with both the scattered $e^+$ and $e^-$ detected 
contribute only about 10\% of the total cross-section,
by comparing the acceptance with single or both 
{\color{OliveGreen} $e^+$ and $e^-$} selected.
}

With the $\theta^{b.p.\pm}_{acc}$ threshold of 25 mrad in the region 
of $|y|\,{>}\, 25$~mm (at $|z|$ = 1~m), the LumiCal
has a Bhabha acceptance of 78.1 nb requiring both $e^+$ and $e^-$ detected. 
This value is nearly twice the $Z\to qq$ cross section 
{\color{black} of
   $\sigma^0_{had}$ = 41.5~nb at the peak (QED corrected from the measured 30 nb 
  \cite{PR427}), 
which provides  sufficient luminosity measurements 
for the $Z$-boson line shape study. 
}

\begin{table}[t!]
\centering
{ \small
\begin{tabular}{ccc}   
\hline
\hline  
    Triggered             &       {Single $e^-$  }         
                           &       {both $e^-$, $e^+$  }        \\ 
                           &    $\sigma$ (nb) & $\sigma$ (nb)       \\
\hline
  $20\,{<}\, \theta^{b.p.\pm}_{acc}\,{<}\, 80$~mrad      & 217.2     &    140.7                    \\
  .and. $|y|\,{>}\, 20$~mm       & 137.5     &    131.2 \\
\hline  
  $25\,{<}\, \theta^{b.p.\pm}_{acc}\,{<}\, 80$~mrad        & 133.4      &   85.9                   \\
  .and. $|y|\,{>}\, 25$~mm      & 82.2       &  78.1  \\
\hline 
\hline
\end{tabular}
}
\vspace{2mm}
  \caption{ The LumiCal acceptance has been analyzed 
   for the Bhabha scattering electrons generated with the BHLUMI  
   in the $\theta$ range of 10 to 120 mrad.
   Events are selected with one or both $e^-$ and $e^+$ 
   detected in the LumiCal acceptance. 
   The cross-sections are listed for the two $\theta$ thresholds 
   of $20, 25\,{<}\, \theta^{b.p.\pm}_{acc}\,{<}\, 80$~mrad, 
   and the joint rectangles between them 
   (dashed lines plotted in Fig.\ref{fig:bhlumi-b2bxy}).
   Excluding the horizontal areas below the beam pipe 
   ($|y|$ = 20, 25~mm,   at $|z|$ = 1~m), 
   the cross-sections with both $e^-$ and $e^+$ detected are about 10\% lower. 
  \label{tab:bhlumi_xsection} }
 \vspace{-5mm} 
\end{table}

\begin{figure}[t!] 
  \centering
  
  \vspace{-6mm}   
  \includegraphics[width=.85\linewidth]{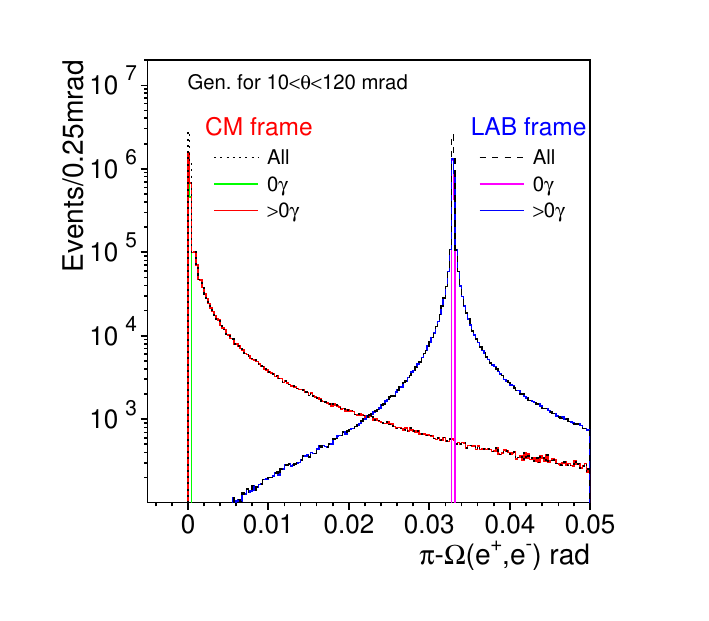}
  \vspace{-4mm}
  \caption{ \label{fig:bhlumi-b2bOA}  
  {\color{black}
  The distributions of acollinearity angles of scattered $e^+, e^-$ 
  generated with the BHLUMI }
  in the CM frame and boosted in the laboratory frame.
  Events with radiative photons cause a deviation from the back-to-back
  acollinearity angle of $e^+$ and $e^-$ $(\Omega(e^+,e^-))$.
  }
  \vspace{4mm}
  \centering
  \includegraphics[width=.70\linewidth]{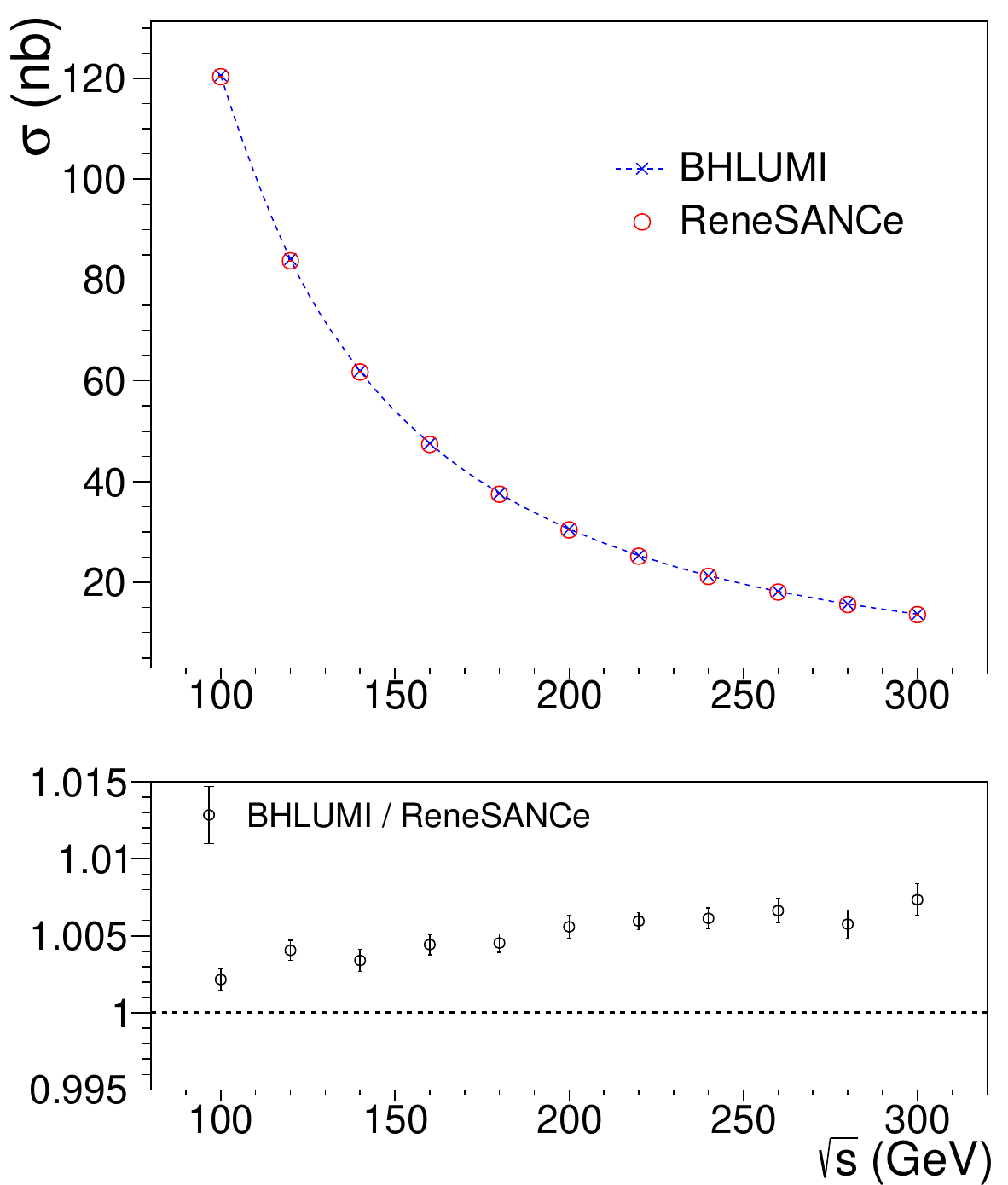}
  
  \caption{ \label{fig:bhlumi-rene}  \color{black}
  Bhabha cross-sections as a function of $\sqrt{s}$
  are plotted for the BHLUMI and ReneSANce simulations
  in the $\theta_{\rm CM}$ range of 30 mrad to $\pi/2$.
  The ratios of them are also plotted.
  }
\end{figure}

\section*{Detecting radiative Bhabha }
\label{sec:RadBhabha}

Bhabha events are selected with 
$e^+$ and $e^-$ back-to-back in directions.
A dominant uncertainty is caused by the production of radiative
photons, resulting in deviation of $e^\pm$ directions.
The distributions of $e^ +$ and $e^-$ 
{\color{black} acollinearity angles,}
generated with the BHLUMI,
are plotted in Fig.~\ref{fig:bhlumi-b2bOA}.

The BHLUMI is reported for a theoretical precision of 0.037\% \cite{Bhlumi20}.
The higher-order calculations have taken into account 
photonic $\mathcal{O} (\alpha^2 L)$
and hadronic $\Delta \alpha_{had}$ corrections. 
The cross-sections calculated at various $\sqrt{s}$ are compared with
the recently developed ReneSANCe \cite{ReneSANCe-v1} 
for the $\theta$ range of 30~mrad to $\pi/2$.
Fig.~\ref{fig:bhlumi-rene} illustrates the results and their ratios.
{\color{black} 
  With the default parameters of the program, the discrepancy 
  is observed at the 0.5~\% level 
  for the wide $\theta$ range. 
  Further investigation in the small-angle region is necessary.
}

The separation of electrons and photons 
is planned using Si-detectors and $2\, X_0$ LYSO arrays
within the LumiCal, with the EM shower deposits in the
$3 \,{\times}\, 3$~mm$^2$ segments of LYSO bars.
The 3~mm pitch of the LYSO bars provides a spatial resolution of
{\color{black} 9~mrad} for two adjacent shower centers 
{\color{black} separated by more than two segmentation units. }

\begin{figure}[b!]  
  \centering
    \vspace{-4mm}
  \includegraphics[width=.85\linewidth]{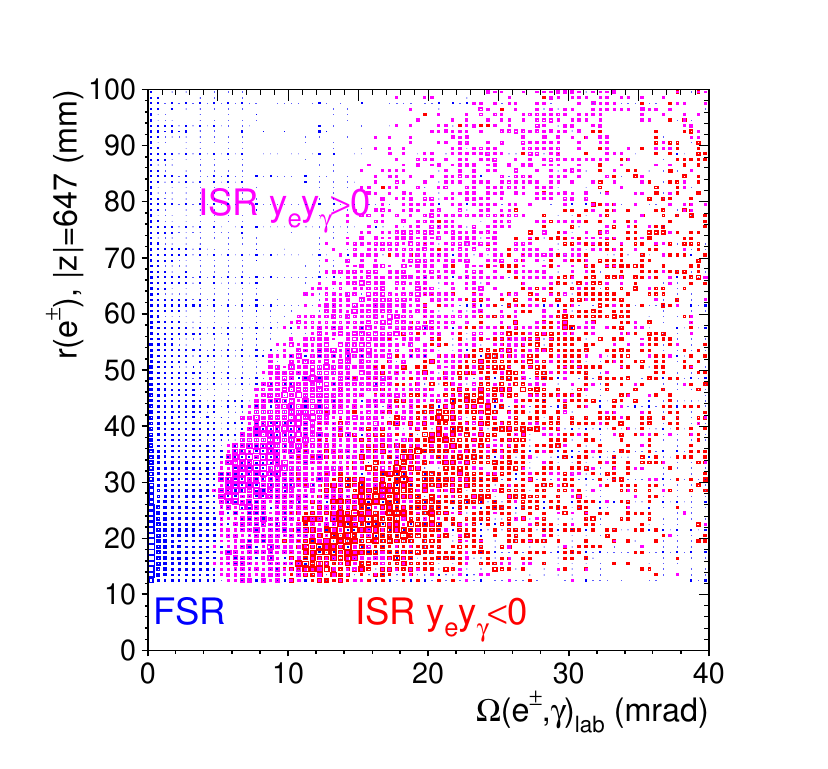}
  
  \vspace{-2mm}
  \caption{ \label{fig:radbha_isrfsr}  
  {\color{black}
  BHLUMI events with radiative photon ($E_\gamma \,{>}\, 0.1$~GeV)
  are boosted for the beam-crossing of 33 mrad.
  The projected positions of $e^\pm$ and $\gamma$ are selected for 
  the LYSO acceptance, of $|y| \,{>}\, 12$~mm, at $|z|$ = 647~mm.
  }
  The $e^\pm$ with $\gamma$ in the same $z$-hemisphere are
  plotted for the  $e^\pm$ radius to $z$-axis versus the $e^\pm,\gamma$ 
  opening angle.
  The photons are identified for FSR (blue)  or
  ISR with the $e^\pm$ separated on the same (red) or opposite (magenta) $y$-sides.
  }
\end{figure}

\begin{table}[b!]  
{ \centering \small 
\begin{tabular}{c|ccc}  
\hline                 
\hline                 
 \multicolumn{4}{r}  
             { $E_\gamma \,{>}\, 0.1$ GeV;  \;\;  \;       
               $ |y(e^+)|, |y(e^-)|,|y(\gamma)| \,{>}\, 12$ mm;   }\\
 \multicolumn{4}{r}                 
             { and           
               $\theta^{b.p.\pm}_{acc}(e^+), 
               \theta^{b.p.\pm}_{acc}(e^-),
               \theta^{b.p.\pm}_{acc}(\gamma) \,{<}\, $80 mrad} \\   
\hline
    {$e^\pm$,{ISR}}
 &  {$e^\pm$,{FSR}}          
 & $\Omega(e^\pm,\gamma) \,{>}\, 9$~mR & ${>}\,5$~mR   \\
\hline  
                 0.91\% & 13.9\% & 1.7\% &  2.6\%  \\
\hline
\hline
\end{tabular}
}
   \vspace{4mm}
  \caption{ {\color{black}
   Events of BHLUMI at $\sqrt{s}$ = 92.3~GeV are boosted for the beam crossing and are
   selected with $e^+$, $e^-$ and $\gamma$ in LumiCal
   of $|y|\,{>}\, 12$ mm at $|z|$ = 647~mm and $\theta^{b.p.\pm}_{acc} \,{<}\, 80$~mrad,
   where $\theta^{b.p.\pm}_{acc}$ are measured with respect to the beam directions. 
   In each $z$-hemisphere, the event fractions 
   having photons of $E(\gamma)\,{>}\, 0.1$~GeV are listed.  
   The fractions with FSR photons of $e^\pm$, $\gamma$ opening angle 
   ($\Omega(e^\pm,\gamma)$)
   larger than 9 mrad and 5 mrad are also listed. }
   The statistical error is 0.1\%.
  \label{tab:radbhlumi} }
\end{table}

The BHLUMI events of $\sqrt{s}$ = 92.3~GeV have
the photons generated with an energy threshold of 5~MeV.
The number of photons 
generated with the YFS exponentiation adheres to a Poisson distribution.
In each event, the photons are categorized as either initial (ISR) 
or final-state radiation (FSR),
with the opening angles in the CM frame being closer to the 
incident or scattered electrons,  
respectively.

Scattered electrons accompanied by 
{\color{black} one or more photons} in the same 
{\color{black} $z$-hemisphere }
account for 40\% of the cases,
with the most energetic photon identified as 
an ISR (20\%) or an FSR (19\%).
Fig.~\ref{fig:radbha_isrfsr} illustrates
the boosted positions of $e^\pm$ (radius to $z$-axis) on the LYSO surface
versus the opening angle with the most energetic photon of $E(\gamma)\,{>}\, 0.1$~GeV
in the same $z$-hemispheres.
Events are selected with all $e^+$, $e^-$, and $\gamma$ 
falling in the LYSO acceptance of $|y|\,{>}\, 12$~mm at $z$ = 647~mm.

In Fig.~\ref{fig:radbha_isrfsr}, the distribution for electrons with FSR 
(blue) reveals a peak near the beam pipe with an opening angle approaching zero.
The ISR photons are  
{\color{black} dispersed around  the outgoing beam pipe 
  of the colliding $e^+$ and $e^-$. }
Electrons with ISR photons are 
distributed in two distinct bands,  
where the $e^\pm$ and $\gamma$ are either on the same side
of the $y$-axis or on opposite sides.

Table~\ref{tab:radbhlumi} lists the event fractions associated with the inclusive 
Bhabha selection for $e^+$ and $e^-$ in the LYSO acceptance range of $|y| \,{>}\, 12$~mm.
Photons are required for $E_\gamma \,{>}\, 0.1$~GeV and are detected in $|y|\,{>}\, 12$~mm.
The event fraction with an ISR, having a large opening angle to the electron
in the same $z$-hemisphere, is 0.91~\%. 
Most of the FSR photons fall within an opening angle of 5~mrad.
{\color{black} 
Electrons associated with an FSR, separated by the LYSO pitches 
assumed for 9 and 5~mrad, account for 1.7 and 2.6~\%, respectively.
}

\section*{LumiCal simulation }
\label{sec:lumi_design}

The Si-detectors in LumiCal detect  
impact positions of charged $e^+$ and $e^-$. The deviation
caused by multiple scattering in the upstream beam pipe materials 
is estimated with the GEANT simulations.
At an incident angle of 25 mrad, the traversing path 
length through a 1~mm thick beam pipe reaches 40~mm.
The corresponding radiation length is $0.11\, X_0$ for a 
Be pipe, and increases to $0.45\, X_0$ for an 
Al pipe.

To achieve a low-mass beam pipe design,
the flat panels joining the Al semicircular beam pipes
($|x| \!<\!6$~mm at $|z|$ = 560~mm, Fig~\ref{fig:MDI-LumiCal3D}) 
 are designed with 1~mm thick Be plates.
The Al pipes are assumed to consist of 1~mm layers.
Electrons of BHLUMI generated at $\sqrt{s}$ = 92.3~GeV
are injected into the GEANT simulations to estimate the effects  
of multiple scattering observed on the Si-wafers at $|z|$ = 560~mm.
The deviations in $\theta$ after traversing the 1 mm thick
Be  or Al beam pipes are fitted to Gaussian functions.
The widths in $\theta$ are 
\begin{eqnarray}  
  52\pm 0.6 \;\mu\mbox{rad} & \;\; \mbox{$|x| \,{<}\, 6$~mm  (1 mm Be)}, \nonumber \\  
  90\pm 0.6 \;\mu\mbox{rad} & \;\; \mbox{$|x| \,{<}\, 6$~mm  (1 mm Al)}. \nonumber     
\end{eqnarray}
The electron hits are populated mostly 
at the edges of $|y|$ = 12~mm.
The errors of 0.6 $\mu$rad are statistical,
corresponding to 12k (29k) events collected in the regions 
of $|x|\!<\!6$~mm ($|x|\!>\!6$~mm), respectively.   

The CEPC colliding beams have bunch sizes of  
$(\sigma_x, \sigma_y, \sigma_z)$ = (6~$\mu$m, 35 nm, 9 mm) \cite{CEPC_TDR}.
The beam-crossing angle of 33~mrad results in an overlap of  
$\sigma_z$ = 380~$\mu$m at IP, {\color{black}
 which implies a smearing of 17~$\mu$rad
 for a track at 25 mrad projected to $|z|$ = 560~mm.
 With the IP distributions randomized for the bunch 
 sizes, the widths of multiple scattering increase by 5\%.
}

Bhabha events are selected with both  
 back-to-back $e^+$ and $e^-$ 
within the detector fiducial edges. 
The spread caused by multiple scattering 
at the edges is analyzed with electrons of BHLUMI. 
{\color{black}
  The $x$, $y$ positions of $e^\pm$ hits are simulated 
  with Gaussian convolutions of 100~$\mu$m in width to the $e^\pm$ positions
  at $|z|$ = 1~m, which 
  corresponds to an angle of 100 $\mu$rad
  {\color{black} for multiple scattering}.    
}

\begin{figure}[t!]
  \centering 
  \includegraphics[width=.78\linewidth]{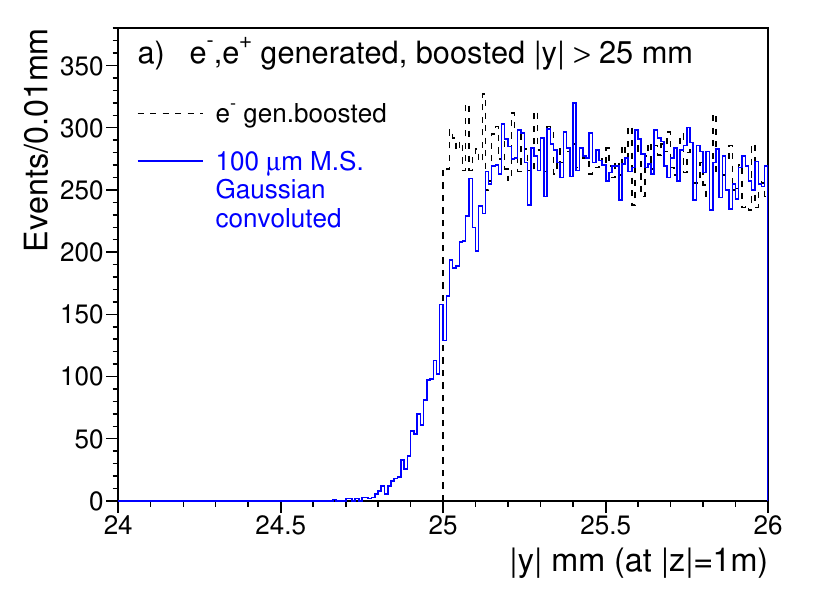}     
  \includegraphics[width=.78\linewidth]{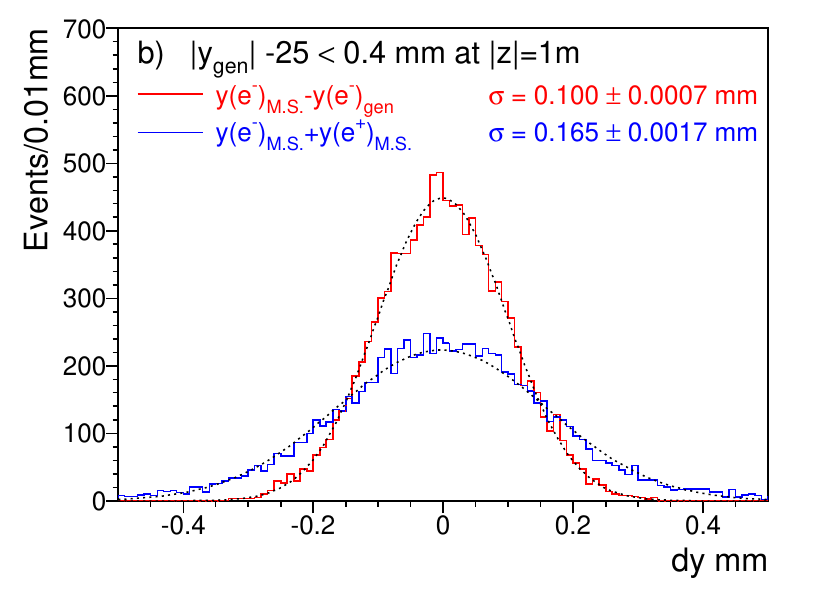}     

  \caption{\label{fig:bhlumi-ms}
  The Bhabha events of BHLUMI, being boosted for the 33 mrad beam crossing,
  are selected for the fiducial acceptance of $|y|\,{>}\, 25$~mm at $|z|$ = 1~m.
  In a), the $e^-$ selected (black dashed) are convoluted with a Gaussian 
  function of {\color{black}
  100~$\mu$m in width on the $e^\pm$ positions}
  to simulate multiple scattering (MS).
  In b), the widths of MS are examined. 
  The differences in $y$ are plotted 
  for the $e^-$ before and after MS convolution. {\color{black}
  The sum in $y$ of the $e^+$ and $e^-$ is also plotted, 
  which represents the projected width in $y$ of 
  the MS smeared acollinearity angle.}
  }
\end{figure}

The distributions of electron impact positions are plotted 
in Fig.~\ref{fig:bhlumi-ms}.a.
The electrons of BHLUMI are selected with $|y|\,{>}\, 25$~mm 
(black dashed line). The Gaussian convolution for multiple scattering
results in a spread in a region of 200 $\mu$m (blue line).         
The differences are  
plotted in Fig.~\ref{fig:bhlumi-ms}.b. The width of
the deviation on $e^-$ positions is reproduced to be
{\color{black} $100\,\pm\,0.7$~$\mu$m. } 
The sum of $e^+$ and $e^-$ positions in $y$ is also plotted.
{\color{black}
The width obtained, $165\pm 1.7$~$\mu$m, is attributed to multiple
scattering and radiative photons. 
If the events are selected with no photons,
the width is reduced to $141\pm1.7$~$\mu$m, 		
which corresponds to the convolution of multiple scattering
with a constant acollinearity 
(Fig~\ref{fig:bhlumi-b2bOA}) of the scattered $e^+$ and $e^-$.
}

\begin{figure}[t!]   
  \centering   
  \includegraphics[width=.78\linewidth]{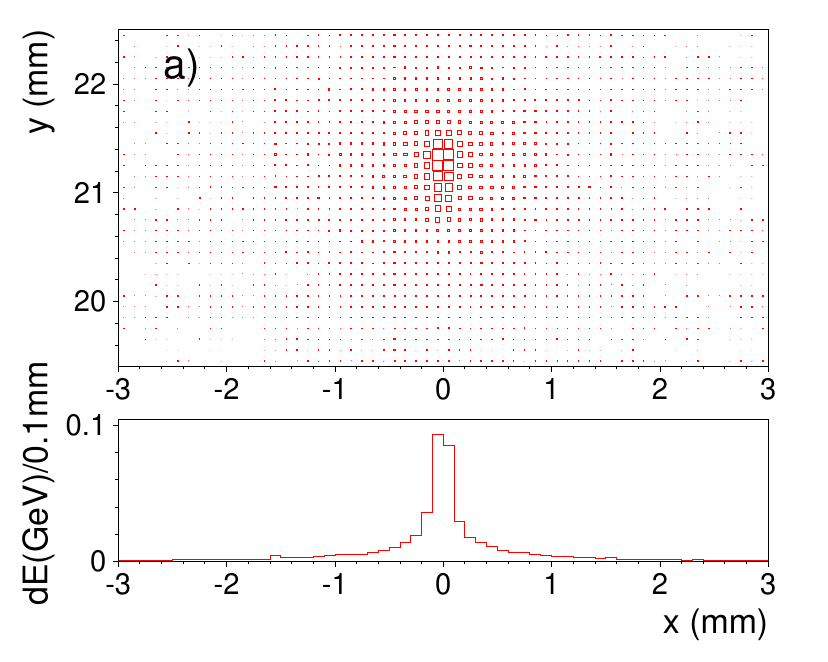}
  \includegraphics[width=.78\linewidth]{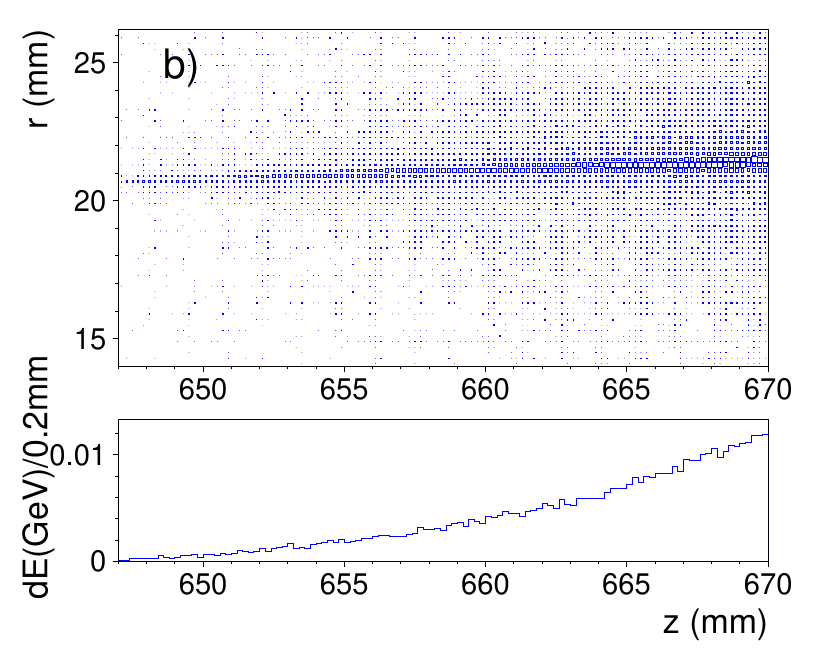}
  
  \vspace{-2mm}
  \caption{\label{fig:gem3_shower}
  Shower profiles of 50 GeV electrons in the LumiCal $2\, X_0$ LYSO crystal
  are shown for the a) $x$-$y$, and b) $r$-$z$ distributions.
  The corresponding projected energy depositions ($dE/dx$) are plotted 
  in histograms.
  The simulations were performed for electrons set at  
  $\theta$ = 32~mrad, vertically at $\phi$ = $90^\circ$
  to the IP and traveling past a 1~mm Be beam pipe.
  The integrated energy deposition ($dE/dz$) along the $z$-axis 
  through the $2\, X_0$ LYSO
  crystal accounts for 1.1~\% of the electron's total energy. 
  }
\end{figure}

\section*{LYSO crystals before flanges}
\label{sec:LYSO_front}


{\color{black} 
Before the beam pipe flanges, 
the LumiCal features Si-detectors and $2\, X_0$ LYSO arrays 
to detect electrons and photons 
in radiative Bhabha events.}
The shower deposits in the LYSO arrays are simulated using GEANT.

The shower profiles of 50~GeV electrons
are modeled at a fixed angle of $\theta$ = 32~mrad, aligned vertically 
with the $y$-axis, after traversing a 1~mm thick Be pipe.
The energy deposits are plotted for the $x$-$y$ and $r$-$z$ distributions
in Fig.~\ref{fig:gem3_shower}.a and b, respectively. 
The shower fragments are narrowly distributed within the 3 mm segmentation width 
of the LYSO array, with approximately  1.1 \% of the 50 GeV electron energy
being deposited in the $2\, X_0$ LYSO crystals.

{\color{black}
To study the separation of electrons from radiative photons,
events of BHLUMI generated at $\sqrt{s}$ = 92.3~GeV are used.
The scattered $e^+$, $e^-$, and photons are injected into the GEANT simulation of
LumiCal.} 
Within the $2\, X_0$  LYSO acceptance region of $|y|\,{>}\, 12$~mm at $|z|$ = 647~mm, 
13.9~\% of the electrons in  
each $z$-hemisphere are accompanied by an FSR photon of $E_\gamma \,{>}\, 0.1$~GeV 
(Table~\ref{tab:radbhlumi}). 
 
\begin{figure}[t!]  
  \centering
  \vspace{-6mm}
  \includegraphics[width=.850\linewidth]{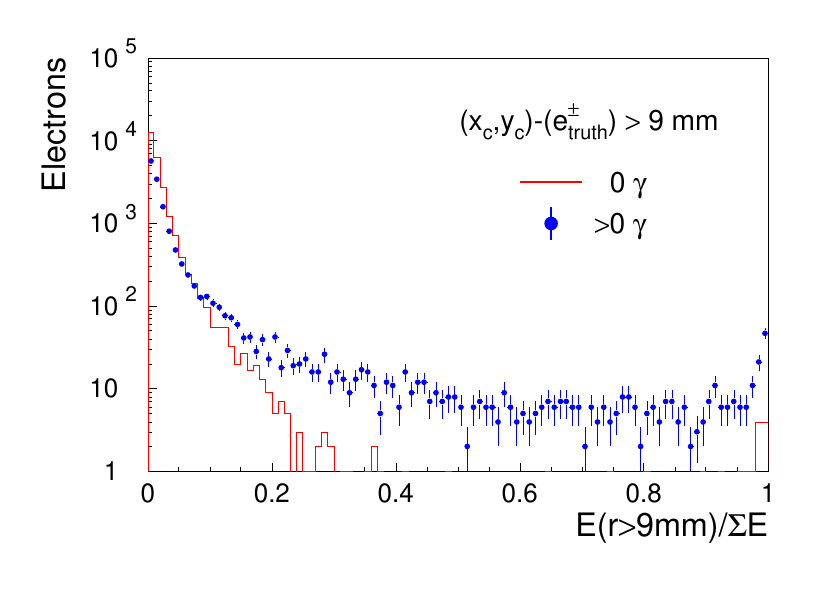} 
  \vspace{-4mm}
  \caption{\label{fig:lumical_bhphot}{ \color{black}
  BHLUMI generated $e^+$, $e^-$, and $\gamma$ are injected to the 
  $2\, X_0$ LYSO at $|z|$ = 647~mm and $|y|\,{>}\, 12$~mm of GEANT simulation. 
  The fractions of external energy deposits in LYSO bars 9 mm away 
  from the $e^\pm$ projected positions, $E(r{>}9{\rm{mm}})/ \Sigma E$, 
  with and without photons in the same $z$-hemisphere.
  Events with photons of $E(\gamma)\,{>}\, 0.1$~GeV are attributed to the
  higher fraction region.
  }}
\end{figure}

{\color{black}
Fig.~\ref{fig:lumical_bhphot} illustrates the fractions of external energy deposits 
in the LYSO cells that are 9~mm away ($E(r{>}9{\rm{mm}})/ \Sigma E$) 
from the incident positions of the $e^\pm$,
for events with and without accompanying photons in the same $z$-hemisphere.   
The showers of 46~GeV electrons  are mostly contained 
within a $3{\times} 3$ LYSO array, \color{black}
with less than 20~\% of the energy deposits occurring outside this region
 (red line of $e^\pm$ without photons). 
 
 Electrons with FSRs are closely attached in most cases, which is  
 seen by the distributions of blue marks in Fig.~\ref{fig:lumical_bhphot}
 with the fraction of external 
 energy peaking at zero.
 The search for FSRs can be conducted 
 by looking for energy deposits in cells that are separated 
 from the electron clusters, which correspond to events with
 $E(r{>}9{\rm{mm}})/ \Sigma E \,{>}\, 0.2$.

 Further optimization of the LYSO crystal dimension is necessary 
 to enhance the efficiency of electron and photon identification.
}

\section*{LYSO crystals behind Bellows}
\label{sec:LYSO_shower}

Bhabha scattering electrons carrying beam energy  {\color{black}
  can be distinguished}
from the beam background by the shower deposits in the calorimeters.
The MDI region has accommodated the LYSO arrays of 150~mm in length ($13\, X_0$) 
between the bellows and the quadrupole magnets.
In front of the long LYSO,
the flange (30~mm) and bellow (46~mm) are made of a total of $4.3 \, X_0$ steel. 
The total amount of upstream materials, including the front $2\, X_0$ LYSO,
adds up to $6.3\, X_0$.

\begin{figure}[t!]  
  \centering
  \vspace{-2mm}
  \includegraphics[width=.720\linewidth]{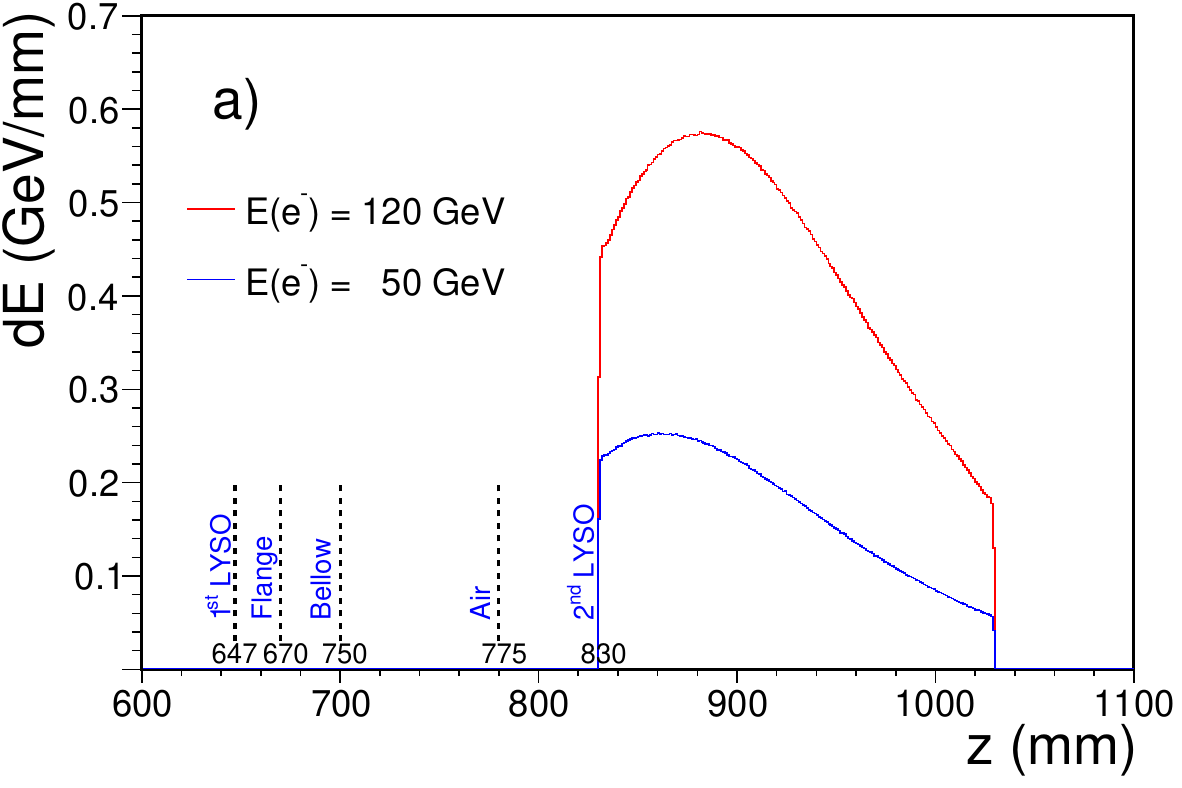}
  \includegraphics[width=.720\linewidth]{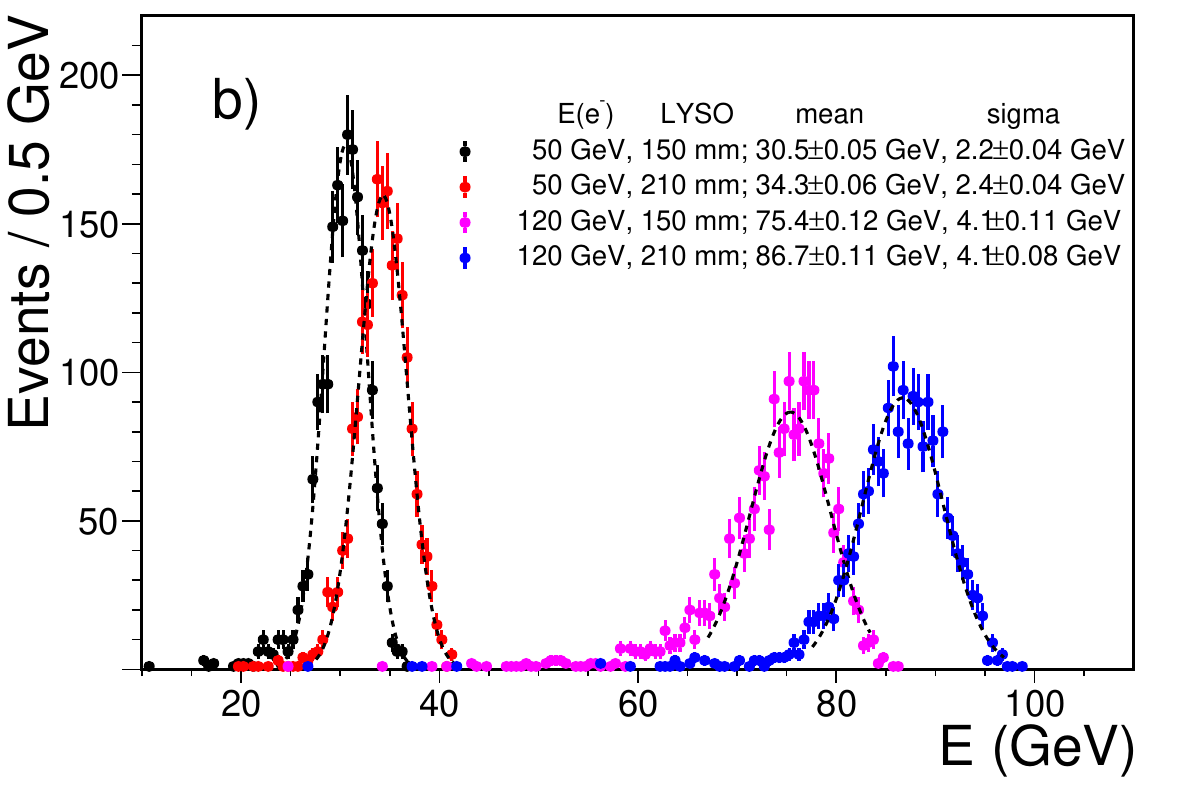}
  \includegraphics[width=.720\linewidth]{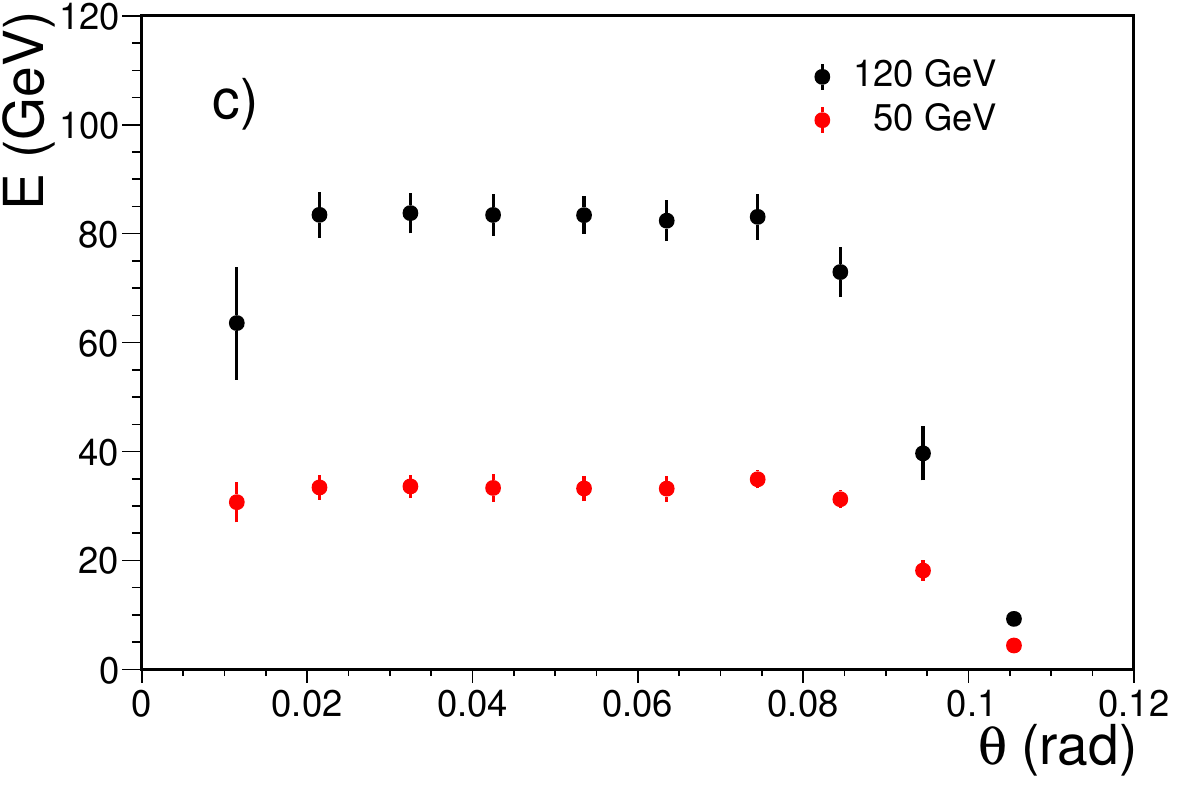}
  \caption{\label{fig:LYSO-reso} {\color{black}
   The long LYSO bars positioned behind the bellows are simulated with lengths 
   of 150~mm and 210~mm.
   The shower profiles are simulated for
   incident electrons at $\theta$ = 44~mrad and $\phi$ = $90^\circ$, with 
   1k events for each beam energy and LYSO configuration.
   The average longitudinal energy deposits in $z$ are plotted in a). 
   The Gaussian fits to the sum $dE/dz$ are plotted in b). 
   For the LYSO of 150 (210)~mm, the means obtained are 30.5 (75.4)~GeV, 
   with the widths of 2.2 (4.1)~GeV for 50 (120)~GeV electrons,
   respectively. 
   The longer 210~mm LYSO collects more $dE/dz$,
   with the means of about 12~\% (15~\%)
   higher than those of 150~mm LYSO.  
   The shower containment with the 210~mm LYSO is plotted in c), 
   with the means and error bars of sum $dE/dz$ for theta values 
   of the LumiCal coverage.
  }}
\end{figure}

The shower profiles are simulated for 50 and 120 GeV electrons 
at a fixed angle of $\theta$ = 44~mrad and $\phi$ = $90^\circ$ 
centered in the LYSO arrays.
The lateral shower deposits along the length of the crystals are plotted 
in Fig.~\ref{fig:LYSO-reso}.a.
The positions of the shower maxima are distributed around 8 to $10\, X_0$.

The energy resolution is compared for the LYSO lengths of 150~mm and 210~mm, 
which are plotted in Fig.~\ref{fig:LYSO-reso}.b.
{\color{black}
  The Gaussian fits to the energy deposits yield 
  the mean values of 30.5~GeV (75.4~GeV) and the widths 
  of 2.2~GeV (4.1~GeV) in 150~mm LYSO, for 50~GeV (120~GeV) electrons,
  respectively.
  With 210~mm LYSO, the mean values increase by 12.5~\% (15.5~\%) 
  for 50~GeV (120~GeV) electrons, respectively, and the widths are
  compatible with those of 150~mm LYSO. 
  }

The shower containment versus the electron incident  $\theta$ angle
is investigated with electrons injected between 
10 and 110~mrad and $\phi=90^\circ$.
In Fig.~\ref{fig:LYSO-reso}.c, the distributions are plotted
for the means and widths of Gaussian fits at each $\theta$ angle.
A plateau for shower containment is noticeable from approximately
20~mrad to 80~mrad. 

{\color{black}
The short length of the LYSO, being preceded by a material thickness 
of $6.3\, X_0$, raises concerns about distinguishing low-energy beam backgrounds.
To enhance the selection of Bhabha events, additional criteria 
   may be introduced by requiring the
   back-to-back coincidence of $e^+$ and $e^-$, 
   with each one matching the front LYSO crystals
   and hits on the Si-detectors.
}

\subsection*{Precision on LumiCal $\theta$ acceptance}
\label{sec:systematics}

The precision on luminosity measurement depends on 
the accuracy of counting Bhabha events in
the LumiCal fiducial area.
{\color{black}
  The detector modules are mounted on the beam-pipe with the
  reference positions fixed to the beam axes on the flanges.}
The survey precision of detector positions 
relative to the IP is the dominant systematic uncertainty.
The technical considerations are discussed in the following.

\begin{figure}[t!]
  \centering
  \includegraphics[width=0.95\linewidth]{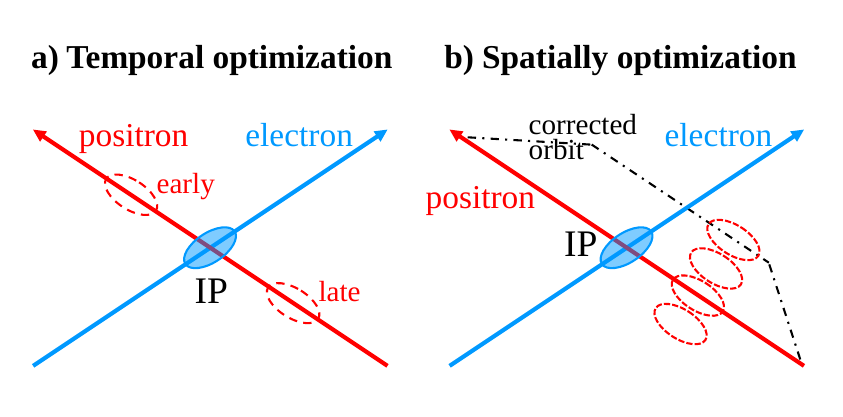}
  \vspace{-4mm}
  \caption{\label{fig:bpm-offset}
  Beam steering and measurements with BPMs are illustrated.
  a) Temporal optimization is conducted with the BPMs in flanges
  on the timing of the crossing bunches at IP.
  b) The transverse positions of the electron and positron beams
  in each beam are measured by the BPMs on each side of the IP,
  with the orbit being deviated (dashed line) by corrector magnets
  to be centered within the beam-pipe. 
  }
\end{figure}

\begin{description}     
\setlength{\itemindent}{-3pt} 
\setlength{\labelsep}{-4pt} 
\setlength{\itemsep}{10pt}  

\item[ \textbf{\textit{a.}} ] 
\textbf{\textit{ MDI LumiCal survey }} \hfill

The LumiCal design has the acceptance region 
of $|y|\,{>}\, 12$~mm on the front Si-wafers at $|z|$ = 560~mm. 
The spread of electron hits, caused by multiple 
scattering through the beam pipe,
results in a symmetrical distribution with a width of approximately 100~$\mu$rad. 
Si-detectors with a resolution of 5~$\mu$m can measure this distribution
across the fiducial edges. The errors of the mean on fiducial edges 
are expected for sub-micron precision. 

The assembly of LumiCal modules 
can be surveyed with sub-micron precision 
{\color{black} using inductive displacement sensors\footnote{
 e.g. industrial products of Micro-Epsilon eddyNCDT.} 
 or a spectral interferometer\footnote{
 e.g. industrial products of Keyence SI-F series.}.
}
 Position monitoring is crucial for the LumiCal fiducial edges.
To achieve a precision of $10^{-4}$ for integrated 
luminosity, the errors on the means of fiducial edges
must be kept to less than 0.6~$\mu$m in the $x$ and $y$ directions, 
and equivalently 25~$\mu$m in the $z$ direction (Eq.~\ref{eq:dLL}).

\item[ \textbf{\textit{b.}} ] 
\textbf{\textit{ Beam Position Monitoring }} \hfill  

Electrons are measured relative to the IP distribution, 
which is ideally the centered position of the beam crossing.
The colliding beams are monitored using   
Beam Position Monitors (BPMs) to track the positions of beam currents
  \cite{BPM_MST33,BPM_NST33}. 
In the MDI region, the BPMs are implemented inside the flanges
and surrounding the single beam pipes before the quadrupole magnets 
(shown in Fig.~\ref{fig:MDI-LumiCal3D}).

The beam steering in real-time employs closed-loop feedback 
to maintain optimal collision parameters.
The BPMs located in the flanges measure coupling
signals generated by the passing of electron and positron bunches, 
which are used to locate the longitudinal positions of the beams. 
The BPMs around each beam pipe are 
used to measure the transverse position of the beams.

  \begin{description}     
  \setlength{\itemindent}{-3pt} 
  \setlength{\labelsep}{-4pt} 
  \setlength{\itemsep}{10pt}  
  
  \item[ \textbf{\textit{b.1}}  ] 
  \textbf{\textit{ Beam Longitudinal Position }} \hfill  
  
  To ensure that the electron and positron bunches arrive 
  simultaneously at the IP, their longitudinal positions 
  must be measured and adjusted 
  by modifying the radio frequency (RF) phase 
  in the electron and positron rings, 
  which alter their arrival time at the IP,
  as depicted in Fig.~\ref{fig:bpm-offset}.a. 
  
  The BPMs in the flanges at $|z|$ = 685~mm measure the relative 
  timing between the electron and positron bunches to determine
  their longitudinal positions.
  Assuming that the colliding bunches 
  pass the BPMs in the flanges at $t_{e^-}$ and $t_{e^+}$ on each side.
  To arrive at the IP simultaneously, the time intervals between the
  crossing bunches shall be exactly
  $\Delta t$ = $t_{e^+} \,{-}\, t_{e^-} = 
  2 \,{\times}\, 685 \;\mathrm{mm} / (3 \,{\times}\, 10^8 \;\mathrm{m/s})$
  = 4.566~ns.
  
  The IP offset in $z$ is required for $\Delta z \,{<}\, 25$~$\mu$m,  {\color{black}
  which corresponds to 1 $\mu$rad on the LumiCal acceptance edge.
  The error on the mean of $\Delta t$ shall be \color{black}
  better than 0.1~ps.}
  
  \item[ \textbf{\textit{b.2}} ] 
  \textbf{\textit{ Beam Transverse Position }} \hfill  
    
  Ideally, the IP position should be centered within the beam pipe at coordinates 
  {\color{black}
  $(x,y)$ = (0,0). The tilted $e+$ and $e^-$ beams having the same offsets in $x$
  would be crossing at $z$ = 0.}
  A beam-beam scan (BBS) technique is utilized to monitor and measure
  deviations in the transverse beam position at the IP. 
  The beam orbits are tuned with corrector magnets near the IP,
  and the sets of four BPMs positioned around each beam pipe at 
  $|z|$ = 1000~mm are used to measure the beam transverse position.
  
  In Fig.~\ref{fig:bpm-offset}.b, a "bump" 
  (represented by a dashed line) demonstrates how 
  corrector magnets adjust the orbit in one ring. {\color{black}
  By scanning the amplitude of this bump, the center of IP distribution
  is optimized for maximum luminosity.
  By steering the $x$ positions of the colliding
  beams, the centering of IP at $z$ = 0 is located.}
 
  \end{description}
\end{description}


\subsection*{Summary}
\label{sec:lumi_summary}

The design of LumiCal in the CEPC MDI region
is evaluated for its ability to detect Bhabha scattering events,
which provide the most accurate reference for determining the 
luminosity of colliding beams.
Achieving the required precision of 10$^{-4}$ at the 
Z-pole energy will enhance the accuracy of Standard Model 
measurements by an order of magnitude.

The detector acceptance for Bhabha scattering events 
is estimated with the BHLUMI generator.
The combination of Si-detectors and LYSO calorimeters
is aimed at identifying $e/ \gamma$ for the measurement of radiative 
photons. 
To minimize the effects of multiple scattering induced by the beam pipe,
a low-mass Be window is utilized.
The primary source of systematic uncertainty will stem 
from the surveying and monitoring of detector positions, 
as well as the beam steering using BPMs
to achieve sub-micron precision at the interaction point.
The precision that is achievable on the LumiCal acceptance positions 
will render the goal for the luminosity measurement.


\section*{Acknowledgments}

The authors would like to thank the assistance of the 
CEPC accelerator group of the Institute of High Energy Physics.
This work has been partially supported by the Institute of Physics, Academia Sinica.

\section*{Declarations}
{\bf Conflict of interest} On behalf of all authors, the corresponding author states that there is no conflict of interest.

\vspace{-2mm}
{}


\begin{thebibliography}{}


\bibitem{CEPC_TDR}
J. Gao et al.,
``CEPC Technical Design Report: Accelerator'',
Radiat Detect Technol Methods,
8 (2024) 1.

\bibitem{Jadach:1996is}
S. Jadach et al.,
``Upgrade of the Monte Carlo program BHLUMI for Bhabha scattering at low angles to version 4.04'',
Comput. Phys. Commun. 102 (1997) 229.


\bibitem{Jadach:1991by}
S. Jadach et al.,
``Monte Carlo program BHLUMI-2.01 for Bhabha scattering at low angles with Yennie-Frautschi-Suura exponentiation'',
Comput. Phys. Commun., 70 (1992) 305.
 
\bibitem{YFS}
D.R. Yennie,  S.C. Frautschi and H. Suura,
``The infrared divergence phenomena and high-energy processes'',
Annals of Physics, 13 (1961) 379.


\bibitem{GEANT3}
R. Brun et al.,
``GEANT Detector Description and Simulation Tool'',
CERN-W5013, (1994).

\bibitem{GEANT4}
S. Agostinelli et al.",
``Geant4—a simulation toolkit'',
Nucl. Instrum. Methods A 506 (2003) 250.

\bibitem{Si-PDG}	
S. Navas et al., ``Review of Particle Physics'', Phys. Rev. D 110, 030001 (2024);
(``35 Particle detectors at accelerators'').

\bibitem{Hou}
Y.H Chang et al., ``A Study of the charge cluster characteristics 
and spatial resolution of a silicon microstrip detector'',
Nucl. Instr. and Meth. A, 363 (1995) 538.

\bibitem{PR427}
ALEPH Collab. et al.",
``Precision Electroweak Measurements on the Z Resonance'',
Physics Reports, 427 (2006) 257-454

\bibitem{Bhlumi20}
P. Janot and S. Jadach,
``Improved Bhabha cross section at LEP and the number of light neutrino species'',
Phys. Lett. B 803 (2020) 135319.


\bibitem{ReneSANCe-v1}
R. Sadykov and V. Yermolchyk,
``Polarized NLO EW $e^+e^-$ cross section calculations with ReneSANCe-v1.0.0'',
Comput. Phys. Commun., 256 (2020) 107445.


\bibitem{BPM_MST33}
J. He et al., ``Beam position monitor design for the High Energy Photon Source'',
Meas. Sci. Technol. 33, 115106 (2022). 

\bibitem{BPM_NST33}
J. He  et al., ``Design and fabrication of button-style beam position monitors 
for the HEPS synchrotron light facility'', 
Nucl. Sci. and Tech. 33, 141 (2022). 





\end{thebibliography}
\end{document}